\documentclass[12pt,a4paper]{article}

\usepackage{setspace}
\usepackage{epsfig,amsbsy,calc,latexsym,psfrag,graphics}
\usepackage{amsmath, amssymb}
\usepackage{enumerate} \usepackage{indentfirst}
\usepackage{graphicx,rotating}

\def\ie{\textit{i.e. }}
\def\eg{{\it e.g.}\ }

\def\etal{{\it et al.}}

\def\be{\begin{equation}}
\def\ee{\end{equation}}
\def\ben{\begin{equation*}}
\def\een{\end{equation*}}
\def\bea{\begin{eqnarray}}
\def\eea{\end{eqnarray}}
\def\bean{\begin{eqnarray*}}
\def\eean{\end{eqnarray*}}
\def\bear{\begin{array}{l}}
\def\eear{\end{array}}

\newcommand{\D}{\mathcal{D}}
\newcommand{\N}{\mathcal{N}}
\newcommand{\Z}{\mathcal{Z}}

\newcommand{\F}{\mathcal{F}}
\def\one{\hbox{1\kern-.8mm l}}
\newcommand{\de}{\partial}  
\newcommand{\nn}{\nonumber\\}   
\newcommand{\ex}[1]{\e^{#1}\!}  
\newcommand{\lin}{\Lambda_{\N\ug1}} 
\newcommand{\www}[1]{\widehat{#1}}  
\newcommand{\vev}[1]{\langle #1 \rangle} 
\newcommand{\bvev}[1]{\biggl\langle #1 \biggr\rangle} 
\newcommand{\prodl}[1]{\!\!\!\!\!\!\!\prod_{#1}\!\!\!\!\!\!\!} 
 
\newcommand{\sdot}{\!\cdot\!}
\newcommand{\virg}[1]{``#1''}   
\newcommand{\ov}[1]{\overline{#1}}  
\def\ug{\!=\!}  
\newcommand{\pat}[1]{\biggl(#1\biggr)}
\newcommand{\paq}[1]{\biggl[#1\biggr]}
\newcommand{\pag}[1]{\biggl\{#1\biggr\}}
\newcommand{\ppt}[1]{\Bigl(#1\Bigr)}
\newcommand{\ppq}[1]{\Bigl[#1\Bigr]}
\newcommand{\ppg}[1]{\Bigl\{#1\Bigr\}}
\DeclareMathOperator{\const}{const}
\DeclareMathOperator{\e}{e}
\DeclareMathOperator{\tr}{tr}
\DeclareMathOperator{\hc}{h\,c}
\DeclareMathOperator{\du}{d}
\newcommand{\di}[2]{\du^{#1}{\!#2}}
\newcommand{\dsl}[2]{\du^{\!\!\!{\scriptscriptstyle-}#1}{\!#2}}
\newcommand{\intg}{\!\int\!\!}    
\newcommand{\fint}[1]{\intg\D#1}  
 
\newcommand{\xtint}{\intg\di{4}{x}\di{2}{\theta}}
\newcommand{\xttint}{\intg\di{4}{x}\di{2}{\theta}\di{2}{\overline{\theta}}}
\newcommand{\pint}{\intg\dsl{4}{p}}
\newcommand{\pintf}{\intg\frac{\dsl{4}{p}}{(p^2+\mu)^2}}  
\newcommand{\ptint}{\intg\dsl{4}{p}\di{2}{\pi}} 
\newcommand{\f}[1]{\Phi_{#1}\,\!}   
\newcommand{\fb}[1]{\ov{\Phi}_{#1}\,\!}
\newcommand{\fv}[1]{\vec{\Phi}_{#1}\,\!}
\newcommand{\fbv}[1]{\ov{\vec{\Phi}}_{#1}\,\!} 
\newcommand{\fp}[1]{\Phi_{#1}^*\,\!}    
\newcommand{\fa}[1]{\Phi'_{#1}\,\!}     
\newcommand{\faF}[1]{\Phi'^{*}_{#1}\,\!}  
\newcommand{\fav}[1]{\vec{\Phi}'_{#1}\,\!}  
\newcommand{\fap}[1]{\vec{\Phi}'^{*}_{#1}\,\!}

\def\hN{\hat{N}}

\begin{document}
\begin{titlepage}
\date{}
\vspace{-2.0cm}
\title{
{\vspace*{0mm} 
\begin{flushright}
\end{flushright}
\vspace{10mm}
}
$\N=1^*$ model superpotential revisited\\(IR behaviour of $\N=4$ limit)
 \vspace*{0mm}}
\author{S.\ Arnone$^{\dagger}$, G.\ Di Segni$^{\dagger}$,
 M.\ Siccardi$^{\dagger,\ddagger}$, and K.\ Yoshida$^{\dagger,\ddagger}$ \\[5mm]
  {\small $^\dagger$ Dipartimento di Fisica, Universit\`a di  Roma
   ``{\it La Sapienza}''} \\
  {\small P.le Aldo Moro, 2 - 00185 Roma, Italy}\\
  {\small and}\\
  {\small $^\ddagger$ INFN, Sezione di Roma I}\\
  {\small P.le Aldo Moro, 2 - 00185 Roma, Italy}\\
  }
\maketitle
\vspace*{0mm}

\abstract {The one-loop contribution to the superpotential, in particular the 
Veneziano-Yankielowicz potential in $\N=1$ supersymmetric Yang-Mills 
model is discussed from an elementary field theory method and the
matrix model point of view. Both approaches are based on the Renormalization 
Group variation of the superconformal $\N=4$ supersymmetric Yang-Mills model.}

\vspace{1truecm}
\vspace{2mm} \vfill \hrule width 3.cm
\begin{flushleft}
e-mail: stefano.arnone@roma1.infn.it\\
e-mail: gabriele.disegni@roma1.infn.it\\
e-mail: matteo.siccardi@roma1.infn.it\\
e-mail: kensuke.yoshida@roma1.infn.it
\end{flushleft}

\end{titlepage}
\vfill
\newpage

\section{Introduction}\label{chap_1}

Several attempts have been made to derive from
dynamical principles the superpotential of $\N=1$ SUSY gauge models
including the celebrated Veneziano-Yankielowicz (VY) potential for the
pure $\N=1$ Super Yang-Mills (SYM) model~\cite{1}.

Within the QFT framework one can distinguish two approaches \ie either
to apply the generalized Konishi anomaly and the corresponding
anomalous Ward-Takahashi identity (AWTI)~\cite{2} or to try to compute
the potential directly from the microscopic Lagrangian with an appropriate 
regularization scheme.

The former approach leads to the formal equivalence \cite{2} with the
highly successful Matrix Model method due to Dijkgraaf and Vafa~(DV)~\cite{3}, 
originally proposed as a spin-off from string theory, whereas in the latter, 
one generally makes appeal to the instanton calculus \cite{4a} so as to take 
into account \virg{non perturbative} effects like the VY potential.

The instanton method is known to produce some ambiguities in certain
cases \cite{4b} although, in the celebrated Seiberg-Witten model, 
agreement with the instanton method is considered as a proof of
correctness of both methods.

On the other hand, there are also a few attempts to compute low energy
quantities like the superpotential by making use of elementary
diagrammatic methods which exploit the covariant supersymmetric
Feynman rule~\cite{5} previously developed for the purpose of the
\virg{perturbative} derivation of the DV correspondence.

This latter gives an efficient way to extract information about
the low energy holomorphic quantities (\virg{F terms}).

The impression here is that one may obtain the superpotential of the
system made up of gluons and some additional \virg{matter} but that it is
difficult to deal with pure gluonic systems~\cite{6}.

In \cite{7}, a diagrammatic derivation of VY superpotential for the pure 
$\N=1$ SYM model has been attempted; the authors have limited themselves 
to the case of the $SU(2)$ gauge group.

The central observation in \cite{7} is that the superpotential of the
superconformal $\N=4$ SYM model in four dimensions is \virg{trivial} in the 
sense that it receives no contributions from non-trivial \textbf{holomorphic}
terms \cite{7b}. The $\N=4$ SYM model is assumed to be UV finite and so is its
mass deformed version ($\N=1^*$ model) \cite{8}. Superconformal
symmetry of the undeformed $\N=4$ SYM implies $\beta(g)=0$; therefore the gauge
coupling constant does not vary with the energy scale.

Applying a Renormalization Group (RG) -type argument~\cite{9} one may try to
compute the (holomorphic) $\N=1$ potential as the difference from the trivial 
$\N=4$
superpotential.

In \cite{7}, one has obtained the VY superpotential for $SU(2)$ pure $\N=1$ 
SYM.

However, there are some ambiguities in the computations presented in
\cite{7}.

First of all, the insufficient analysis of anticommuting external
field ($W_\alpha$ or $\lambda_\alpha$) makes the generalization to
$N_c>2$ difficult.

Moreover, the limiting procedure to evaluate the \virg{difference} of
potentials
$$\Delta V= V_{\N=1}-V_{\N=4}$$
and reduce it to the VY form as the regularizing mass parameter $\mu$
flows from large $M_0$ (corresponding to $\N=1$ SYM) to $M\sim 0$
(corresponding to $\N=4$ SYM) lacks of mathematical rigour.

Indeed at the end of computation, the one-loop potential takes
the form \be\label{eff_act_a} \const W^2\log\Biggl\{
\frac{1+\overline{\alpha}g_0^2(W^2/M^3)}{1+\overline{\alpha}g_0^2\bigl[W^2/(M_0M^2)\bigr]}\Biggr\},
\ee with $\overline{\alpha}$ being a numerical constant.

The VY form of the potential can appear if one assumes
$\overline{\alpha}g_0^2(W^2/M^3)\gg 1\gg
\overline{\alpha}g_0^2\bigl[W^2/(M_0M^2)\bigr]$ and, hence, approximate it
with \be \const W^2\log\Bigl(\overline{\alpha}g_0^2(W^2/M^3)\Bigr), \ee
On the other hand, a naive IR limit $M\to 0$ makes the whole
potential logarithmically divergent. 

Moreover, in order to arrive at the quoted result, eq. (\ref{eff_act_a}), 
the authors~\cite{7} have adopted the Gaussian approximation \ie the effective
coupling obtained in the intermediate stage of computation has been
truncated beyond the quadratic term.

As we see later, the justification for such \virg{Gaussian
approximation} also depends on the smallness of
$\overline{\alpha}g_0^2\bigl[W^2/(M_0M^2)\bigr]$.

The same RG-type approach (with respect to $\N=4$
SYM model) has been used with complete success in Matrix Model 
computation~\cite{10}. In this paper, the authors have applied the
Matrix Model method by Dijkgraaf and Vafa~\cite{3} to the same model
parametrized by the \virg{floating} mass $\mu$ ($M\leq\mu\leq M_0$).

The crucial point here is again to appeal to the assumed triviality of the
holomorphic part of the $\N=4$ superpotential, corresponding to the $M\to0$
limit. In this way, the authors of \cite{10} were able to uniquely determine
the overall coefficient $C_{\hN}$ for the measure of the matrix
integral.

In particular, their computation shows that the leading term as
$\hN\to\infty$ exhibits a \textbf{smooth} limit as $M\to0$. Thus one can 
define $C_{\hN}$ in
such a way that the $M\to0$ limit gives the required boundary
condition without any ambiguity. It was shown that this
definition of the matrix integral yields the complete $\N=1^*$
superpotential including the VY term as well as the perturbative
corrections.

Adding the symmetry breaking potential at the beginning, one can also 
successfully deal with a spontaneously broken $\N=2$ SYM model, 
obtaining the perturbative version of Seiberg-Witten solution \cite{11}.

In this approach, moreover, one does not appear to encounter any of the 
ambiguities plaguing the instantonic computation of VY potential.

It should be emphasized that in the Matrix Model computation, 
the use of the same idea of triviality of the $\N=4$ superpotential 
does not lead to any IR divergence so long as one is
interested in leading $\hN$ results.

The paper is organized as follows: section (\ref{chap_2}) is devoted to 
retracing the results of \cite{7} and \cite{10} and their reanalysis. In section 
(\ref{chap_3}) we will attempt to construct a more convincing diagrammatic 
computation of the $SU(N_c)$ SYM superpotential from the QFT point of view, 
while section (\ref{chap_4}) will contain our conclusions.

\section{QFT and Matrix Model derivation of
Superpotential}\label{chap_2}

In this section, we will briefly review and compare the methods of \cite{7} and
\cite{10}.

\subsection{ERG approach to $\N=1$ Superpotential}\label{chap_2_qft}

In \cite{7}, one starts with the microscopic action for the $\N=1^*$ model
with gauge group $G=SU(N_c)$.  \bea\label{s}
S_{\N=1^*}(V,\f{i},\fb{i};g_0)&=&
\frac{1}{16}\xtint\frac{1}{\hat{g}_0^2}W^2+\hc \nonumber\\
&&+2N_c\xttint\sum_{i=1}^3\fb{i}\ex{g_0V}\f{i} \nonumber\\
&&+\frac{\imath g_0}{\sqrt{2}}\xtint
f_{abc}\frac{\epsilon_{ijk}}{3!}\f{i}^a\f{j}^b\f{k}^c+\hc \nonumber\\
&&+\frac{1}{2}\xtint\sum_{i=1}^3M_0^i\f{i}^2+\hc, \eea where
$\frac{1}{\hat{g}_0^2}=\frac{1}{g_0^2}+\frac{\imath\vartheta}{8\pi^2}$
(canonical representation). Note that in the original presentation
\cite{7} the authors have used the so-called holomorphic
representation while here we will be using the
canonical representation.

For large $M_0^i$ ($\equiv M_0$), this model can be regarded as a
$\N=1$ SYM model, regularized by a mass deformed $\N=4$ SYM \cite{9}.

It is believed that the model is free of UV divergences for an
arbitrary set of masses ($M_0^i$) just as the original
superconformal model without any mass deformation~\cite{8}.

The pure $\N=1$ SYM can be realized in the limit $M_0^i=M_0\to\infty$
and $g_0\to0$ with
\be\lin=\frac{M_0}{g_0^{2/3}}\exp-\frac{8\pi^2}{3N_cg_0^2}\ee  held
fixed.

On the other hand, the $M_0^i=M\to0$ limit at fixed $g_0$ should realize 
the $\N=4$ SYM for any $g_0$.

In \cite{7} the following three-stage procedure has been used to
obtain the superpotential.

\subsubsection{The holomorphic reduction}\label{hol}

Following \cite{5}, one can approximately integrate out the
antichiral components $(\fb{i})_{i=1,2}$ (we regard
$\fv{3}$ and $\fbv{3}$ as external at this stage), thus obtaining the 
effective $\fv{1},\fv{2}$ action written in momentum space\footnote{Following 
the lead in~\cite{5}, we will Fourier 
transform all superspace coordinates}:
\be\label{eff_act_mom}
\frac{1}{2}\ptint\f{ia}^*(p,\pi)\bigl(-p^2+\pi_\alpha\www{W}^\alpha+\www{A}(\f{3})+M_0\bigr)_{ia,jb}
\f{jb}^*(-p,-\pi),\ee \bean
\text{where}&&\www{W}^\alpha\to\www{W}^\alpha_{ia,jb}=W_c^\alpha
F^c_{ab}\delta_{ij},\\
&&\www{A}\to\www{A}_{ia,jb}=\frac{g_0}{\sqrt{2}}\f{3c}F^c_{ab}\epsilon_{ij}\equiv\frac{g_0}{\sqrt{2}}(\vec{\Phi}\cdot\vec{F})\otimes\imath\sigma_2.
\eean Note that in writing (\ref{eff_act_mom}), the chiral field
$\fv{3}$ too is treated as if it were constant. However, it is easy to
see that, for large $\overline{M}_0$, only the lowest frequency
components of $\fv{3}$ contribute when integrated with respect to
$\fv{1}$ and $\fv{2}$. Eq. (\ref{eff_act_mom}) gives the
effective holomorphic propagator of $\fv{1}$ and $\fv{2}$ (valid only
for the evaluation of low energy amplitudes).

\subsubsection{Exact Renormalization Group reduction.}\label{erg}

In general, one can transform a path integral with an action like 
(\ref{eff_act_mom}) into another of similar form, where the regularizing 
parameter $M_0$ has been changed to $M<M_0$. This is equivalent to 
K.~Wilson's decimation
method in lattice models~\cite{9b} and to the variation
of the cutoff in continuum QFT~\cite{9c}.  The simple formula for implementing 
this change is Zinn-Justin's transformation~\cite{12}:
\bea\label{zj}\fint{\Phi}\exp\biggl[-\frac{1}{2}\int\fp{}(-p)\frac{1}{D_1(p)+D_2(p)}\fp{}(p)-V(\Phi)\biggr]=\nonumber\\
\fint{\f{}}\D\fa{}\exp\biggl[-\frac{1}{2}\int\fp{}(-p)D_1^{-1}(p)\fp{}(p)+\nonumber\\
-\frac{1}{2}\int\fp{}(-p)D_2^{-1}(p)\fap{}(p)-V(\f{}+\fa{})\biggr]
\eea where $D_i(p)$'s are the regulated propagators.

In our case, the situation is somewhat simpler as the effective
action, eq. (\ref{eff_act_mom}), is quadratic in $\fv{1,2}$. Then the
required equivalent of eq. (\ref{zj}) is:

\bea\label{quadeffact}
&&\int \prod_{i=1,2}\D \fv{i} \exp\frac{\imath}{2}\ptint
\fp{ia}[-p^2+\pi_\alpha\www{W}^\alpha+\www{A}(\f{3})+M_0]_{ia,jb}\fp{jb}\nonumber\\
&=&\int \prod_{i=1,2}\D \fv{i}\D\fav{i}\exp\Bigl[\frac{\imath}{2}\ptint
\fp{ia}[-p^2+\pi_\alpha\www{W}^\alpha+\www{A}(\f{3})+M]_{ia,jb}\fp{jb}+\nonumber\\
&&+\frac{\imath}{2}\intg\faF{ia}\bigl[\bigl(-p^2+\pi_\alpha\www{W}^\alpha+\www{A}(\f{3})+M_0\bigr)
\bigl(-p^2+\pi_\alpha\www{W}^\alpha+\www{A}(\f{3})+M\bigr)^{-1}\bigr]_{ia,jb}\faF{jb}\Bigr]
\nonumber\\ \eea In eq. (\ref{quadeffact}) 
$\fv{}$ and $\fav{}$ have been diagonalised in order to cancel the mixed product in (\ref{zj}).

As explained in~\cite{7}, the first term in RHS, with reduced mass $M$, will reproduce, 
in the vanishing $M$ limit, the amplitude for the $\N=4$ SYM,
while the second term should contribute the non-trivial part of the
\virg{Wilsonian action} $S_{M_0}-S_{\N=4}$.

The gaussian integral over $\fa{1,2}$ has been exactly computed in
\cite{7} in the case of $N_c=2$. We only quote the result: 
\bea\label{eff-coupling}
&&\frac{W_1W_2}{8\pi^2}\Biggl\{\log\biggl(\frac{M}{M_0}\biggr)^2
+\log\biggl(\frac{1+\phi^2/M^2}{1+\phi^2/M_0^2}\biggr)+\nonumber\\
&&+2\frac{(\phi_1^2+\phi_2^2)}{M^2}\biggl(\frac{M}{\phi}\biggr)^3
\biggl[\tan^{-1}\biggl(\frac{\phi}{M}\biggr)-\biggl(\frac{\phi}{M}\biggr)\biggr]
-(M\leftrightarrow M_0)\Biggr\}+\nonumber\\ &&-(W_1\leftrightarrow W_2)
\eea where $W_\alpha=W_\alpha^3$ (diagonal component),
$(\fv{3})^2\equiv\vec{\phi}^2=\phi_1^2+\phi_2^2+\phi_3^2$.

\subsubsection{Integration over $\f{3}$} 

In order to arrive at the
superpotential as a function of the gluon supermultiplet only, one must
integrate over $\fv{3}\equiv\vec{\phi}$, with the effective coupling
given in eq.~(\ref{eff-coupling}).

To this end, one may try to apply the same procedure used for
$\D\fv{1}\D\fv{2}$, \ie first integrate out $\fbv{3}$ and then apply the
ERG transformation, $M_0\to M$. Dealing with a non-quadratic action, 
eq.~(\ref{eff-coupling}), one must in principle apply Zinn-Justin's formula
(\ref{zj}) in its unsimplified form. However, in \cite{7}, a Taylor expansion 
up to the second order was performed first and, then, the previous procedure 
to the resulting quadratic action for $\fv{3}$ was applied.

The quadratic approximation to (\ref{eff-coupling}) is \ben \frac{4
W_1
W_2}{16\pi^2}\biggl[\log\biggl(\frac{M}{M_0}\biggr)^2+\biggl(\frac{\phi}{M}\biggr)^2-\frac{2}{3}\frac{\phi_1^2+\phi_2^2}{M^2}\biggr]-(W_1\leftrightarrow
W_2) \een
(omitting $M_0^{-2}$ terms) and the result after $\D\fv{3}$
 integration is \be\label{sup_2}
\exp\frac{\imath}{4\cdot 16\pi^2}\int
W^2\Biggl\{\log\biggl(\frac{M}{M_0}\biggr)
+\log\Biggl(\frac{1+\frac{g_0^2W^2}{32\pi^2\cdot 3
M^3}}{1+\frac{g_0^2W^2}{32\pi^2\cdot 3 M_0M^2}}\Biggr)\Biggr\}.\ee

If one adds the $\log(M/M_0)^2$ contribution from (\ref{eff-coupling}),
and the gauge kinematical term in $S_{\N=4}$, eq.~(\ref{sup_2}) takes 
the form \be
\exp\frac{2\imath}{128\pi^2}\int
W^2\Biggl\{\log\biggl(\frac{M}{\Lambda}\biggr)^3
+\log\Biggl(\frac{1+\frac{g_0^2W^2}{32\pi^2\cdot 3
M^3}}{1+\frac{g_0^2W^2}{32\pi^2\cdot 3
M_0M^2}}\Biggr)+\frac{\imath\vartheta_0}{2}\Biggr\}.\ee

To conclude that the potential is of VY type, in the vanishing $M/M_0$ limit, 
one must be able to ascertain 
\be \label{estimate_su2}
\log\biggl(\frac{1+\alpha g_0^2W^2/M^3}{1+\alpha
g_0^2W^2/(M_0M^2)}\biggr) \sim\log\biggl(\frac{\alpha
g_0^2W^2}{M^3}\biggr) \ee

If one can justify this assumption, then the superpotential for $\N=1$
SYM (with $N_c=2$) takes the VY form \be\label{vy-su2} W_\text{eff}=
\frac{1}{128\pi^2}\int\biggl\{2\log\biggl(\frac{S}{3\Lambda^3\cdot32\pi^2}\biggr)+\imath\vartheta_0\biggr\}S
\qquad (S\equiv W^2)\ee If we look for the extrema of (\ref{vy-su2}),
we find \be\label{vev_su2} \langle W^2/(32\pi^2) \rangle \sim (\pm
\exp(-\imath\vartheta_0/2)\Lambda'^3. \ee where \be
\Lambda'=\left(\frac{3}{\e}\right)^{1/3}\Lambda \ee Note that
no use of instantons has been made to obtain the results
(\ref{vy-su2})-(\ref{vev_su2}).

\subsection{Matrix Model method to compute $\N=1$ potential}\label{chap_2_mm}

It has been suggested in~\cite{3} that the large $\hN$ limit of a 
certain Matrix Model can
reproduce the holomorphic superpotential of a wide class of gauge field
theories.

At the beginning, it was believed that such a correspondence was limited
to the perturbative corrections to the superpotential, thus excluding $\N=1$
VY potential \cite{2}. This \virg{inability} was related to the
fact that one could not determine unambiguously the overall
coefficient for the matrix integral measure \cite{10b}.

Kawai and his collaborators, after establishing the direct
correspondence between DV methods and certain generalizations of gauge
field theories (on non-commutative space time) \cite{13}, tried
to use the triviality of $\N=4$ superpotential as a boundary
condition for determining the unknown overall coefficient of the matrix
measure \cite{10}.

In \cite{10}, one starts with a $\hN$-dimensional hermitian Matrix
Model characterized by the tree-level potential \be\label{mm_sm}
S_m=\frac{\hN}{g_m}\tr\biggl(\Phi_1[\Phi_2,\Phi_3]+W(\Phi_1)+\frac{M_2}{2}\Phi_2^2+\frac{M_3}{2}\Phi_3^2\biggr)\ee
and the Dijkgraaf-Vafa-type free energy
\be\Z=\exp\biggl(-\frac{\hN^2}{g_m^2}F_m\biggr)=C_{\hN}\intg\di{}{\Phi_1}\di{}{\Phi_2}\di{}{\Phi_3}\exp(-S_m).\ee

To determine the overall coefficient $C_{\hN}$, one considers the
specific form ($\N=1^*$ model),  \be
S_m=\frac{\hN}{g_m}\tr\biggl(\Phi_1[\Phi_2,\Phi_3]+\frac{M_1}{2}\Phi_1^2+\frac{M_2}{2}\Phi_2^2+\frac{M_3}{2}\Phi_3^2\biggr), \ee
and tries to fix $C_{\hN}$ by demanding that the $\N=4$ SYM limit ($M_i\to0$) 
reproduce the \virg{trivial} model  
\ben \F_{\N=4}=\lim_{M_i\to
0}\F_{\N=1^*}=\frac{\pi\imath\tau_0 g_m^2}{N_c}. \een

Evaluating the matrix integral for small $M_i$, one obtains
\be\label{z_n1*} \Z_{\N=1^*} \approx
C_{\hN}J_{\hN}\biggl(\frac{2\pi}{\hN}\biggr)^{\hN^2} \biggl(\frac{2\pi
g_m}{\hN M_1 M_2 M_3}\biggr)^{\hN/2} \ee where $J_{\hN}$ is the 
result of the integral over the \virg{angular variables} of a hermitian
matrix\footnote{As is usual,
\begin{equation*}  
  \intg\di{}{\www{\Phi}}=J_{\hN}\intg\du 
  \vec{\lambda} \Delta^2(\lambda), 
\end{equation*} where 
the Van Der Monde determinant for the Gaussian Unitary Ensemble takes the form 
$\Delta^2(\lambda)= \prod_{i<j} (\lambda_i - \lambda_j)^2$.
}.

The remarkable fact about eq.~(\ref{z_n1*}) is that the leading 
term, of order $\hN^2$, is smooth in the $M_i\to 0$ limit (constant), and the IR divergent term
is subleading in $\hN$.

This fact guarantees the successful outcome of Kawai's scheme and the
result is
\be\label{c_n}C_{\hN}=\biggl(\frac{\hN^3g_0^2}{(2\pi)^3\ex{2/3}g_m^2}\biggr)^{\hN^2/2}\ex{-\pi\imath\tau_0\hN^2/N_c}. \ee

Now we can assume that this value, for given ($\hN,g_m$), be valid for any
potential $W(\Phi_1)$ in eq.~(\ref{mm_sm})

As has been shown explicitly in \cite{10}, choosing
$W(\Phi_1)=(M_1/2)\Phi_1^2$ and $M_i\equiv M_0\to\infty$, one can obtain
the superpotential of $\N=1$ SYM following the DV prescription
\cite{3}, \be W_\text{eff}^\text{SYM}
=N_c\frac{\partial}{\partial S }\biggl[\frac{S^2}{2}\log\biggl(\frac{\ex{3/2}\Lambda^3}{S}\biggr)\biggr]
=N_c S \biggl[1-\log\biggl(\frac{S}{\Lambda^3}\biggr)\biggr], \ee which
is none other than the VY potential~\cite{1}.

\subsubsection{$\N=1^*$ models}\label{sec:221}

One can also generalize \cite{10} the above computation to the case
of an arbitrary $W(\Phi_1)$ in eq. (\ref{mm_sm}). In particular, in the
simple case that $W(\Phi_1)=(M_1/2)\Phi_1^2$, one obtains the
superpotential for $\N=1^*$ model which include both the VY term and
perturbative corrections.

The corresponding matrix integral is given by

\be\Z=C_{\hN}\intg\di{}{\Phi_1}\di{}{\Phi_2}\di{}{\Phi_3}\exp\bigl[-S_{\N=1^*}(\Phi_i;M_i)\bigr],\ee
where \be\label{s_n1}\quad
S_{\N=1^*}=\frac{\hN}{g_m}\tr\biggl(g_0\Phi_1[\Phi_2,\Phi_3]
+\frac{1}{2}\sum_{i=1}^3 M_i\Phi_i^2\biggr),\ee and $C_{\hN}$ is as in
(\ref{c_n}).  One is now interested in the small gauge coupling, $g_0$, and
large but finite $M_i$ region.

After integrating over two of the three $\Phi$'s and rescaling the diagonal
elements of the remainder, one can rewrite $\Z_{\N=1^*}$ in the following
form \be\label{z_n1_2}
\Z_{\N=1^*}=\ex{-\pi\imath\tau_0\frac{\hN^2}{N_c}}\gamma^{\frac{\hN^2}{2}}\cdot \int \du \vec{\lambda}\ex{-\frac{1}{2}\sum_{i=1}^{\hN}\lambda_i^2}\prod_{i<j}\biggl\{\frac{(\lambda_i-\lambda_j)^2}{1+\gamma
(\lambda_i-\lambda_j)^2}\biggr\},\ee where $\gamma\equiv g_0^2g_m/(\hN
M_0^3)$ and $M_0\equiv(M_1M_2M_3)^{1/3}$.

One can evaluate (\ref{z_n1_2}), expanding it in terms of the \virg{small}
parameter $\gamma$.

Thus one can compute the DV free energy, $F_m$, as \bea F_m &=&
-\frac{g_m^2}{2}\biggl[\log\biggl(\frac{g_0^2g_m}{M_0^3\hN}\biggr)-\frac{2\pi\imath\tau_0}{N_c}\biggr] +
\nonumber\\
&&-\frac{g_m^2}{\hN^2}\log\intg\du \vec{\lambda}\ex{-\frac{1}{2}\sum\lambda_i^2}
\prod_{i<j}\biggl\{\frac{(\lambda_i-\lambda_j)^2}{1+\gamma
(\lambda_i-\lambda_j)^2}\biggr\}.\eea

In the planar limit, exploiting the DV correspondence ($\hN\to\infty$, $g_m\sim S$)  
\bea
F_m^\text{(planar)}&=&-\frac{S^2}{2}
\biggl[\log\biggl(\frac{g_0^2S}{M_0^3\ex{3/2}}\biggr)-\frac{2\pi\imath\tau_0}{N_c}\biggr]
-\frac{S^2}{\hN^2}\Biggl\{-\gamma\bvev{\sum_{i>j}(\lambda_i-\lambda_j)^2}\biggr.+
\nonumber\\
&&+\frac{\gamma^2}{2}\biggl[-\bvev{\sum_{i>j}(\lambda_i-\lambda_j)^2}^2+
\bvev{\biggl(\sum_{i>j}(\lambda_i-\lambda_j)^2\biggr)^2}\biggr. +\nonumber\\
&&+\biggl.\bvev{\sum_{i>j}(\lambda_i-\lambda_j)^4}\biggr] +\ldots\Biggl\}=\nonumber\\
&=&
-\frac{S^2}{2}\biggl[\log\biggl(\frac{g_0^2S}{M_0^3\ex{3/2}}\biggr)-\frac{2\pi\imath\tau_0}{N_c}\biggr]+
\nonumber\\ &&+S^2\biggl\{-\frac{g_0^2S}{M_0^3}+
\frac{7}{2}\biggl(\frac{g_0^2S}{M_0^3}\biggr)^2-23\biggl(\frac{g_0^2S}{M_0^3}\biggr)^3+\ldots\biggr\},
\eea where \be \bigl\langle\bigl\{\ldots\bigr\}\bigr\rangle
\equiv \displaystyle{\frac{\int \du \vec{\lambda} \Delta^2(\lambda) \ex{-\frac{1}{2}\sum_{i=1}^{\hN}\lambda_i^2}
\bigl\{\ldots\bigr\}}{\int \du \vec{\lambda} \Delta^2(\lambda) \ex{-\frac{1}{2}\sum_{i=1}^{\hN}\lambda_i^2} }}. \ee

The effective potential is given by \bea
W_\text{eff}(S)&=&N_c\frac{\de F_m^\text{(planar)}}{\de
S}=-N_cS\biggl[\log\biggl(\frac{S}{\Lambda^3}\biggr)-1\biggr]+
\nonumber\\
&&-N_cS\biggl[-3\biggl(\frac{g_0^2S}{M_0^3}\biggr)+14\biggl(\frac{g_0^2S}{M_0^3}\biggr)^2
-115\biggl(\frac{g_0^2S}{M_0^3}\biggr)^3+\ldots\biggr] \eea which
agrees with previous results \cite{14}.

\subsubsection{$\N=2$ models}\label{sec:222}

As shown in~\cite{10}, with exactly the same definition of the 
integration measure one can deal with spontaneously broken models.

In this case, one has to add to the tree-level action, eq.~(\ref{s_n1}), a
symmetry breaking term.

In order to analyse the breaking pattern \ben U(N)\to U(N_1)\times
U(N_2)\qquad N_1+N_2=N,\een it suffices to introduce the cubic term
\be\label{delta_W} \Delta
W(\Phi_1)=\epsilon\tr\biggl(\frac{1}{3}\Phi_1^3-v^2\Phi_1\biggr), \ee and take
the $\epsilon \to 0$ limit at the end of computation. Then, one can study the $\N=2^*$ model which can
go over to $\N=2$ SYM (Seiberg-Witten model) in the infinite mass limit.

One sets $M_2=M_3=\Lambda_0\gg M_1\sim 0$ and
$\vev{\Phi}_\text{cl}=\pm v$.

The relevant matrix integral is again \be\label{z_n2}
\Z_{\N=2^*}=C_{\hN}\intg\di{}{\Phi_1}\di{}{\Phi_2}\di{}{\Phi_3}\exp-\bigl[S_{\N=1^*}+\Delta
W(\Phi_1)\bigr].\ee

To study $\N=2$ SYM, for instance, one can go over to the limit that
$\Lambda_0\to\infty$ and integrate out $\Phi_2,\Phi_3$. In this case,
since the cubic term can be neglected with respect to the $\Phi_2,\Phi_3$
mass terms, the only trace of the $\N=4$ regularization is the coefficient
$C_{\hN}$.

Eq. (\ref{z_n2}) is then reduced to the matrix integral with a single
matrix \bea\label{z_n2_2} &&\Z=C_{\hN}\biggl[\frac{2\pi
g_m}{\hN\Lambda_0}\biggr]^{\!\hN^2}\!\intg\di{}{\Phi}
\exp-\frac{\hN}{g_m}\tr\biggl[\epsilon\biggl(\frac{1}{3}\Phi^3-v^2\Phi\biggr)\biggr]\nonumber\\
&&=C_{\hN}J_{\hN}\biggl[\frac{2\pi
g_m}{\hN\Lambda_0}\biggr]^{\!\hN^2}\!\int \du \vec{\phi} \Delta^2(\phi)
\exp-\frac{\hN\epsilon}{g_m}\sum_{i=1}^{\hN}\biggl[\frac{1}{3}\phi_i^3-v^2\phi_i\biggr]\nonumber\\
\eea

Such integral has been already studied \cite{11} except for the
explicit reference to $C_{\hN}$.

On the other hand, in the $\N=4$ approach by Kawai \etal~\cite{10}, the
unambiguous definition of $C_{\hN}$ leads to a very clear interpretation
of dynamical cutoffs appearing in the computation. This is
particularly important when dealing with the non-perturbative
formulation of the SW model where UV divergences appear in the corresponding matrix
integral.

To analyze further the matrix integral (\ref{z_n2_2}) one, as usual,
picks up the classical vacua of the potential (\ref{delta_W}) \be
\vev{\Phi}_\text{cl}=\pm v \ee and constructs the series of vacua, 
characterized by $\hN_1$ eigenvalues at $v$ and $\hN_2$
eigenvalues at $-v$ ($\hN_1+\hN_2=\hN$).

Expanding the action (\ref{z_n2_2}) around the respective classical
vacua \be S'=\frac{\hN}{g_m}\sum_{i=1}^{\hN_1}\biggl[\frac{2\epsilon
v}{2}p_i^2+\frac{\epsilon}{3}p_i^3\biggr]+
\frac{\hN}{g_m}\sum_{j=1}^{\hN_2}\biggl[-\frac{2\epsilon
v}{2}q_j^2+\frac{\epsilon}{3}q_j^3\biggr]\ee and taking account of all
possible ways to choose $\hN_1$ out of $\hN$ eigenvalues, $\Z$ can be
written as \be
\Z= {\hN \choose \hN_1} \Lambda_0^{-\hN^2}\ex{-\frac{\pi\imath\tau_0}{N_c}\hN^2}\!\!
\int \du \vec{p} \du \vec{q} \prod_{i=1}^{\hN_1}\prod_{j=1}^{\hN_2}(2v+p_i+q_j)^2\prodl{1\leq
i<l\leq\hN_1}(p_i-p_l)^2 \prodl{1\leq
j<k\leq\hN_2}(q_j-q_k)^2\ex{-S'}.\ee  Wick-rotating the $\vec{q}$
variables and with a suitable rescaling \bea
\Z&=&{\hN \choose \hN_1} \Lambda_0^{-\hN^2}\ex{-\frac{\pi\imath\tau_0}{N_c}\hN^2}
\frac{(\alpha V)^{2\hN_1\hN_2}}{\alpha^{\hN^2}}\!\!
\int \du \vec{p}\prodl{1\leq
i<l\leq\hN_1}(p_i-p_l)^2\ex{-\sum_m^{\hN_1}\frac{p_m^2}{2}}\times\nonumber\\
&&\times\du \vec{q}\prodl{1\leq
j<k\leq\hN_2}(q_j-q_k)^2\ex{-\sum_n^{\hN_2}\frac{q_n^2}{2}}\times\nonumber\\
&&\times\biggl\{\prod_{i=1}^{\hN_1}\prod_{j=1}^{\hN_2}\biggl[1+\frac{p_i-\imath
q_j}{\beta}\biggr]^2
\ex{-\sum_i^{\hN_1}\frac{p_i^3}{3\beta^3}}\ex{-\sum_j^{\hN_2}\frac{q_j^3}{3\beta^3}}
\biggr\},\eea where $V=(v-(-v))=2v$, $\alpha=(\hN\epsilon V/g_m)^{1/2}$
and $\beta=(V/\alpha)$.

Here, one can consider $(1/\beta)$ as the \virg{small} parameter for
the perturbative expansion of $\Z$ as for the corresponding free
energy $F_m$.

In the planar limit ($\hN\to\infty$ with $g_m\hN_i/\hN\to S_i$)
($i=1,2$) the first few terms of $F_m$ are the following:

\subsubsection*{$\mathbf{0^{th}}$ order}

\be\label{f0}
F_m^{(0)}=\frac{\pi\imath\tau_0}{N_c}(S_1+S_2)^2+(S_1+S_2)^2\log\frac{\Lambda_0}{V}
+\sum_{i=1}^2\frac{S_i^2}{2}\log\biggl(\frac{\ex{3/2}\epsilon
V^3}{S_i}\biggr). \ee

This is essentially equal to the expression found in \cite{10} except for
the fact that there one assumes $\Lambda_0\approx V$, while here $\Lambda_0$ is clearly identified with the
regularization mass from the mass-deformed $\N=4$ SYM models ($\N=2^*$
models).

Thus $\Lambda_0$ is related to the dynamical cutoff for $\N=2$ SYM, \ie
\be \Lambda_{\N=2}/\Lambda_0=\exp[-8\pi^2/(2N_cg_0^2)] \ee and the
kinematical term of the mass deformed $\N=4$ SYM is equal to \be
\pi\imath\tau_0=N\log[\Lambda_{\N=2}/\Lambda_0]. \ee

Eq. (\ref{f0}) can be rewritten in term of the physical quantities for
$\N=2$ SYM, \be F_m^{(0)}=(S_1+S_2)^2\log\frac{\Lambda_{\N=2}}{V}
-\sum_{i=1}^2\frac{S_i^2}{2}\log\biggl(\frac{S_i}{\ex{3/2}\epsilon
V^3}\biggr). \ee

The corresponding superpotential is \bea
W_\text{eff}^{(0)}=\sum_{i=1}^{2}N_i\frac{\de F}{\de
S_i}&=&N_c(S_1+S_2)\log\biggl(\frac{\Lambda_{\N=2}}{V}\biggr)^2
-\sum_{i=1}^2 N_i S_i \biggl[ \log \biggl(\frac{S_i}{\epsilon V^3}\biggr)-1\biggr] \nonumber\\
&\equiv& \sum_i N_i S_i \biggl[\log\biggl(\frac{\Lambda_i^3}{S_i}\biggr)+1\biggr] \eea
The last equality in the above is obtained by introducing the low
energy cutoffs $(\Lambda_i)^2_{i=1,2}$ defined by \be\epsilon
V^3\biggl(\frac{\Lambda_{\N=2}}{V}\biggr)^\frac{2N}{N_i}\equiv\Lambda_i^3\ee
More explicitly \bea
\Lambda_1^{3N_1}=\epsilon^{N_1}V^{N_1-2N_2}\Lambda^{2N_c}_{\N=2},\nonumber\\
\Lambda_2^{3N_2}=\epsilon^{N_1}V^{N_2-2N_1}\Lambda^{2N_c}_{\N=2}, \eea
which agrees with the definition introduced in \cite{17}.

\subsubsection*{Higher order terms in $\mathbf{1/}\boldsymbol{\beta}$}

In order to calculate higher order corrections in $1/\beta$ we have to expand the partition 
function: \be
\Z\sim\bvev{\prod_{i=1}^{\hN_1}\prod_{j=1}^{\hN_2}\biggl[1+\frac{(p_i-\imath
q_j)}{\beta}\biggr]^2
\exp{\biggl(-\sum_{i=1}^{\hN_1}\!\frac{p_i^3}{3\beta}\biggr)}\exp{\biggl(-\sum_{j=1}^{\hN_2}\!\frac{(\imath
q_j)^3}{3\beta}\biggr)}}\ee where \bea\label{bvev}
\bvev{\bigl(\ldots\bigr)}&\equiv&\int \du \vec{p}\prodl{1\leq
i<l\leq\hN_1}(p_i-p_l)^2\exp{\biggl[-\sum_{m=1}^{\hN_1}\frac{p_m^2}{2}\biggr]}\nonumber\\
&&\int \du \vec{q}\prodl{1\leq
j<k\leq\hN_2}(q_j-q_k)^2\exp{\biggl[-\sum_{n=1}^{\hN_2}\frac{q_n^2}{2}\biggr]}\bigl(\ldots\bigr).\eea

Computing $F_m$ to order $1/\beta^2$ and $(1/\beta^2)^2$ respectively, one obtains,
in the planar limit \bea \pat{\frac{1}{\beta^2}}:\quad
F_m^{(1)}(S_1,S_2)&=&-\frac{1}{V^3\epsilon}\paq{\frac{2}{3}S_1^3-\frac{2}{3}S_2^3-5S_1S_2(S_1-S_2)}\nonumber\\
\pat{\frac{1}{\beta^2}}^2:\quad
F_m^{(1)}(S_1,S_2)&=&-\frac{1}{V^6\epsilon}\paq{\frac{8}{3}S_1^4-\frac{8}{3}S_2^4-\frac{91}{3}S_1S_2(S_1^2+S_2^2)+59S_1^2S_2^2}\nonumber\
\eea

These results are essentially equivalent to those
in \cite{11,14}. 

\subsubsection{Non perturbative approach to SW model}

The perturbative treatments in the Matrix Model approach, described in
sections \ref{sec:221} and \ref{sec:222}, can be converted to the
exact (``non perturbative'') results by means of the large $\hN$
matrix technology~\cite{15b}.

The coefficient $C_{\hN}$ of the matrix integral measure introduced
in~\cite{10} is still relevant in this case.

As is well known, in the analytical treatment of the matrix integral
in large-$\hN$ limit, a certain integral of the matrix resolvent $R(x)$
is divergent~\cite{15}, which necessitates the introduction of
cutoffs.

The coefficient $C_{\hN}$, as defined in
section~\ref{chap_2_mm}, gives a simple and consistent
interpretation of such cutoffs in terms of the regularizing mass in the
original mass deformed $\N=4$ SYM model.

We will illustrate this in the case of Seiberg-Witten type
model~\cite{18} with $\N=2$ SUSY and gauge symmetry broken as:
\begin{equation}\label{eq:45}
  SU(N_c) \rightarrow [U(1)]^{N_c-1}.
\end{equation}
Here, we mostly follow the presentation of \cite{19}.

The relevant matrix partition function is given by
\begin{equation}\label{eq:46}
  \Z = e^{-\frac{\hN}{g_m^2} \F_m} = C_{\hN} \int \prod_{i =1}^{ 3}
  \du \Phi_i \exp-[S_{\N=1^*}+W(\Phi_i)].
\end{equation}
Here $S_{\N=1^*}$, given by eq. (\ref{s_n1}), is the tree-level
potential for the mass-deformed $\N=4$ SYM with masses for chiral
fields $\vec{M} = (0, \Lambda_0, \Lambda_0)$, and $W(\Phi_i)$ is the symmetry 
breaking inducing term. One has
\begin{equation}
  W' (x) \propto \prod_{i=1}^{N_c} (x -a_i), \qquad \qquad \qquad
  \sum a_i = 0.
\end{equation}

After integrating out the heavy components $\Phi_2$ and $\Phi_3$
and diagonalizing $\Phi_1$, eq. (\ref{eq:46}) is reduced to
\begin{equation}
  \Z= \frac{1}{(\Lambda_0)^{\hN^2/2}} e^{-\pi \imath \tau_0 \hN^2/N_c}
  \int \du \vec{\phi} \Delta^2(\phi)
  \exp -\frac{\hN}{g_m} \sum_i W(\phi_i).
\end{equation}
Rescaling $\phi_i$ with the regularizing mass $\Lambda_0$, \ie
$\phi_i \equiv \Lambda_0 \lambda_i$, one has
\begin{equation}
  \Z= e^{-\pi \imath \tau_0 \hN^2/N_c} \int \du \vec{\lambda}
  \Delta^2(\lambda) \exp-\frac{\hN}{g_m}
  \sum_i \widetilde{W}(\lambda_i),
\end{equation}
where $\widetilde{W}(\lambda_i)$ is again a polynomial with
$\widetilde{W}'(x) \propto \prod_{i=1}^{N_c} (x-\tilde{a}_i)$.

If we write the large-$\hN$ matrix integral as
\begin{equation}
  \int \du \vec{\lambda} \Delta^2(\phi)
  \exp-\frac{\hN}{g_m} \sum_i
  \widetilde{W}(\lambda_i) \equiv e^{-\frac{\hN^2}{g_m^2}
  \widetilde{\F}_m},
\end{equation}
then the physical matrix free-energy is
\begin{equation}
  \F_m = \frac{\pi \imath \tau_0}{N_c}g_m^2 + \widetilde{\F}_m,
\end{equation}
where the first term comes from the factor $C_{\hN}$, while
$\widetilde{\F}_m$ can be written in terms of the density
$\tilde{\rho}(\lambda)$ of matrix eigenvalues in the large-$\hN$ limit
\begin{equation}\label{eq:rho}
  -\frac{\hN^2}{g_m^2} \widetilde{\F}_m = g_m \sum_i \int_{A_i}
  \textrm{d} \lambda \tilde{\rho}(\lambda) \widetilde{W} (\lambda)
  - g_m^2 \sum_{i,j} \int_{A_i} \textrm{d} \lambda \int_{A_j}
  \textrm{d} \lambda' \tilde{\rho}(\lambda) \tilde{\rho}(\lambda')
  \log |\lambda - \lambda'|.
\end{equation}
$(A_i)_{i=1}^{N_c}$ are the intervals where the eigenvalues are
concentrated and correspond to the cuts on the real axis for the
matrix resolvent
\begin{equation}
  R(x) = g_m \sum_i \int_{A_i} \textrm{d} \lambda
  \frac{\tilde{\rho}(\lambda)}{(x-\lambda)}.
\end{equation}
They are analytically defined by the auxiliary function
\begin{equation}
  \mathcal{Y}^2 = W'^2 + f_{N_c-1} = \prod_{i=1}^{N_c} (x - a_i^+)(
  x - a_i^-),
\end{equation}
which gives the corresponding branch points and $A_i=[a_i^-,
a_i^+]$. $f_{N_c-1}$ is the $(N_c-1)$-th order polynomial defined by
\begin{equation}\label{eq:55}
  f_{N_c-1}=\frac{4g_m}{\hN} \left\langle \tr (\frac{
  \widetilde{W}'(\Phi) - \widetilde{W}'(x)}{x - \Phi} )
  \right\rangle_\Phi.
\end{equation}
As usual, one considers the model in terms of the independent input
parameters $a_1, \ldots, a_{N_c-1}$, $(\sum_{i=1}^{N_c} a_i = 0)$,
and
\begin{equation}
  S_i = \frac{1}{2 \pi i} \oint_{A_i} R(x) \du x, \qquad i=1,
  \ldots N_c \qquad (\sum_{i=1}^{N_c} S_i = g_m).
\end{equation}

To study the variation of $\widetilde{\F}_m$ with respect to the
collective variables $S_i$'s, one needs to consider the structure of
the Riemann surface defined by the behaviour of $R(x)$ with respect
to the cuts $A_i$,
\begin{equation}
  R(x + \imath \epsilon) - R(x- \imath \epsilon) = -2\pi \imath \rho (x), 
  \qquad x \in A'_i \qquad \mbox{(gluing conditions).}
\end{equation}
The above defines the two-sheeted surface connected with  $N_c$ tubes.
$R(x)$ and $\mathcal{Y}(x)$ are single valued functions on this
surface. One can show, from eq. (\ref{eq:rho}), that
\begin{equation}\label{eq:F_m}
  \frac{\partial \widetilde{\F}_m}{\partial S_i} = \frac{1}{2}
  \int_{C_i} \mathcal{Y} \du x + const. = \int_{C_i} R \du x + const.,
\end{equation}
where $C_i$ is the line connecting the points at infinity $P (x_+ =
\Lambda_0' \rightarrow \infty)$ on the first sheet and $Q (x_+ =
\Lambda_0' \rightarrow \infty)$ on the second sheet passing through
the cut $A_i$. One can get rid of the unknown $const.$ by 
rewriting eq. (\ref{eq:F_m}) as
\begin{equation}\label{eq:dFdS}
  \frac{\partial \widetilde{\F}_m}{\partial S_i} - \frac{\partial \widetilde{
  \F}_m}{\partial S_j} = \frac{1}{2} \oint_{B_{i,j}} \mathcal{Y} \du x = 
  \oint_{B_{i,j}} R \du x.
\end{equation}
The closed curve $B_{i,j}$ is given by $B_{i,j} = C_i - C_j$. 
$\Lambda_0'$ represents the U-V cutoff necessary to compute the 
matrix model.

\subsubsection*{Seiberg-Witten theory}

The Seiberg-Witten theory is given by the extremum condition for the 
effective action (DV action)
\begin{eqnarray}
  W_\text{eff} &=& \sum_{h=1}^{N_c} \frac{\partial}{\partial S_h} \F_m \nonumber \\
  &=& \sum_{h=1}^{N_c} \frac{\partial}{\partial S_h} \widetilde{\F}_m + N_c 
  \frac{\partial}{\partial g_m} \left( \frac{\pi i \tau_0}{N_c} g_m^2 \right) 
  \nonumber \\
  &=& 2 \pi \imath \tau_0 \sum_{h=1}^{N_c} S_h + \sum_{h=1}^{N_c} \frac{ \partial
  }{\partial S_h} \F_m
\end{eqnarray}
\ie
\begin{equation}
  \frac{\partial}{\partial S_j} \sum_{h=1}^{N_c}  (2 \pi \imath \tau_0 S_h + \frac{\partial
  \widetilde{\F}_m}{ \partial S_h} ) =0 \qquad j=1,2,\ldots,N_c.
\end{equation}

This can be rewritten as
\begin{eqnarray}
  \left(\frac{\partial }{\partial S_i} - \frac{\partial}{ \partial S_{i+1}} \right) \sum_{ h 
  =1}^{N_c} \frac{\partial \widetilde{\F}}{\partial S_h} &=& 0 \qquad i = 1, \ldots 
  N_c-1, \\
  2 \pi \imath \tau_0 + \frac{\partial}{\partial S_1} \sum_{h=1}^{N_c}  \frac{\partial 
  \widetilde{\F}}{\partial S_h}&=&0.
\end{eqnarray}
From the expression for $\partial \F /\partial S_i$ given above, eq. (\ref{eq:dFdS}), 
these equations reduce to the following equations for the resolvent $R$:
\begin{eqnarray}
  \oint_{B_{i,i+1}} \sum_h \frac{\partial R}{\partial S_h} &=& 0 \label{eq:62a} \\
  \int_{C_1} \sum_h \frac{\partial R}{\partial S_h} + 2 \pi \imath \tau_0 &=& 0. 
  \label{eq:62b}
\end{eqnarray}
The $\tau_0$ term in eq.~(\ref{eq:62b}) comes from the $C_{\hN}$ coefficient 
of the matrix integral measure. The bare coupling constant $\tau_0$ codifies 
the regularization condition of our model in the original QFT form, \ie 
the $\N=2^*$ model with mass $\vec{M} = (0,\Lambda_0,\Lambda_0)$, which 
defines the regularized $\N=2$ SYM in the limit $\Lambda_0 \rightarrow \infty, 
g_0 \rightarrow \infty$, with the $\N = 2$ dynamical cutoff
\begin{equation}
  \Lambda^2_{\N = 2} = \Lambda_0^2 e^{ -8 \pi^2 / g_0^2 N_c}
\end{equation}
kept constant.

Thus, in such regularization scheme
\begin{equation}
  2 \pi \imath \tau_0 = 2 N_c \log \left(\frac{ \Lambda_{\N=2}}{\Lambda_0}
  \right) .
\end{equation}

It has been shown~\cite{24b} that in the particular breaking pattern, 
eq.~(\ref{eq:45}), the function $f_{N_c-1}$, introduced in eq.~(\ref{eq:55}), 
has to reduce to a constant. In more detail, conditions (\ref{eq:62a}, 
\ref{eq:62b}) are solved by the particular choice of inputs (see, 
\eg,~\cite{19}):
\begin{equation}\label{eq:SWcurve}
  \mathcal{Y}^2 = \widetilde{W}'^2 - 4 \Lambda^{2 N_c} \qquad 
  (\Lambda = const.).
\end{equation}
that is the Seiberg-Wittern curve~\cite{18}. Under this choice, for $x 
\rightarrow \infty$
\begin{eqnarray*}
  R \sim \frac{1}{x^{N_c}} &\qquad \qquad & \mbox{on the first sheet,} \\
  R \sim x^{N_c} &\qquad \qquad & \mbox{on the second.}
\end{eqnarray*}

From eq. (\ref{eq:SWcurve}), one can write 
\begin{equation}
  \sum_h \frac{\partial R}{\partial S_h} = - \frac{\du}{\du x} \log R(x)
\end{equation}
from which one can easily appreciate that the condition (\ref{eq:62a}) is satisfied. As for eq.~(\ref{eq:62b}), one has
\begin{equation}
  \int_{C_1} \sum_h \frac{\partial R}{\partial S_h} = -  \log R \Big\rvert_Q^P
\end{equation}
Now
\begin{eqnarray*}
  (\log R)_P &=& \log ( W' - \sqrt{ W'^2 - 4 \Lambda^{2N_c}} ) \Big\rvert_{x \sim 
  \Lambda'_0} \\
  &\sim& \log \{{\Lambda'}_0^{N_c} - {\Lambda'}_0^{N_c}[ 1 - 2 \left(\Lambda / 
  \Lambda'_0 \right)^{2N_c}] \} \\
  &=& \log \left[ 2 \frac{\Lambda^{2N_c}}{{\Lambda'}_0^{N_c}}\right],
\end{eqnarray*}
\begin{eqnarray*}
  (\log R)_Q &=& \log ( W' + \sqrt{ W'^2 - 4 \Lambda^{2N_c}} ) \Big\rvert_{x \sim 
  \Lambda'_0} \\
  &\sim& \log \{{\Lambda'}_0^{N_c} + {\Lambda'}_0^{N_c}[ 1 - \ldots ] \} \\
  &=& \log \left[ 2 {\Lambda'}_0^{N_c}\right],
\end{eqnarray*}
and
\begin{equation}
  \log R \Big\rvert_Q^P = \log \left( \frac{\Lambda}{\Lambda'_0} \right)^{2 N_c}.
\end{equation}
This will be exactly cancelled by the $2 \pi i \tau_0$ term in eq.~(\ref{eq:62b}) 
if
\begin{equation}
  \left\{ 
    \begin{array}{rcl}
      \Lambda & = & \Lambda_{\N =2 } \\
      \Lambda'_0 & = & \Lambda_0
    \end{array}
  \right.
\end{equation}

Thus one can conclude that the SW ansatz is the solution of extremum 
conditions for our matrix model if the cutoffs $\Lambda'_0$ and $\Lambda$ in the 
matrix model computation are identified in terms of the $\N=2^*$ regularization of
$\N =2 $ SYM.

For the rest of the calculation, one can go along the standard SW construction of the 
prepotential~\cite{15}.

\section{An improved QFT derivation of superpotential}\label{chap_3}

In this section, we would like to describe an improved computation of the 
superpotential for the general $SU(N_c)$ SYM model.

We deal with the model given by eq. (\ref{s}) and apply the techniques
outlined in sections \ref{hol} and \ref{erg}, \ie covariant supersymmetric
Feynman rules and \virg{ERG} variation of regularizing mass parameters.

With respect to eq. (1), we have seen in section \ref{chap_2} that
 integrating out $(\bar{\Phi}_{i}^a)_{a=1}^{N_c^2-1}$ ($i=1,2$)
reproduces the holomorphic action (\ref{eff_act_mom}) which is
quadratic in $\f{i}^a$.

In this section, however, we will make use of the fact that the actions to
be dealt are always quadratic and take a shortcut substituting Z-J
transformation (\ref{zj}) with a simple rescaling transformation of the
field variables.

Thus we apply the following transformation written in momentum space
\be \fp{i,a}(p,\pi)\rightarrow\ppg{(-p^2+\pi\www{W}+\www{A}+M)^{1/2}
(-p^2+\pi\www{W}+\www{A}+M_0)^{-1/2}}_{ia,jb}\fp{j,b}(p,\pi) \ee The
corresponding Jacobian is
\bean&&\pag{\det\ppq{(-p^2+\pi\www{W}+\www{A}+M)
(-p^2+\pi\www{W}+\www{A}+M_0)^{-1}}}^{-1/2}\\
&=&\exp-\frac{\imath}{2}\ptint\tr\ppq{\log(p^2+\pi\www{W}+\www{A}+M)-\log
(p^2+\pi\www{W}+\www{A}+M_0)}\eean (the 4-momenta are Wick rotated) where
again $\www{A}(\fv{3})$ is treated as if it were a constant matrix (see section
\ref{chap_2_qft}). 

Note that in the present situation, the Konishi anomaly is the simple
consequence of our holomorphic Feynman rule. \cite{7}

The resulting intermediate effective action (as a functional of $W$
and $\fv{3}$) is given by \be
-\frac{1}{2}\ptint\tr\ppq{\log(\www{\Gamma}+M_0)-\log(\www{\Gamma}+M)}, \ee
with $\www{\Gamma}\equiv p^2+\pi\www{W}+\www{A}$.  
Effecting the fermionic integration yields 
\begin{eqnarray}\label{eff_n}
  &&-\frac{1}{8}\pint\ppg{\tr\ppq{(p^2+\mu+\www{A})^{-1}\www{W}_1(p^2+
  \mu+\www{A})^{-1}\www{W}_2} -(\www{W}_1\leftrightarrow\www{W}_2) 
  }_{\mu=M}^{\mu=M_0}= \nonumber\\
  &=&-\frac{1}{8}\pintf \textrm{tr} \paq{\pat{1+\frac{\www{A}}{p^2+\mu}
  }^{-1}\www{W}_1\pat{1+ \frac{\www{A}}{p^2+\mu}}^{-1}\www{W}_2 -( 
  \www{W}_1\leftrightarrow\www{W}_2)}_{\mu=M}^{\mu=M_0}\nonumber\\
  &\equiv& S^{(1)}(M_0)-S^{(1)}(M).
\end{eqnarray}
The matrices $\www{W}$ and
$\www{A}$ are defined in eq. (\ref{eff_act_mom}).

As in the previous section, we consider the effective action
(\ref{eff_n}) only up to the quadratic term in $\fv{3}$. Then \bean
S^{(1)}(\mu)\sim\pintf\tr_{\text{adj}}
\pag{(\www{W}_1\sdot\www{W}_2)-\frac{g_0^2/8}{(p^2+\mu)^2}
\paq{(F\sdot\phi)^2(F\sdot W_1)(F\sdot W_2)+\\+(F\sdot
W_1)(F\sdot\phi)^2(F\sdot W_2)+ (F\sdot\phi)(F\sdot
W_1)(F\sdot\phi)(F\sdot W_2)}-(W_1\leftrightarrow W_2)}\eean (where it
is understood that the momentum integration should be done only after
taking the difference $S^{(1)}(M_0)-S^{(1)}(M)$.

Making use of the commutation relation between $F\sdot\phi$'s and
$F\sdot W$'s, one can rewrite it as 
\begin{eqnarray}
  S^{(1)}(\mu) &\sim& 2\pintf\tr_{\text{adj}} \pag{(\www{W}_1\sdot\www{W}_2) 
  - \frac{g_0^2/8}{(p^2+\mu)^2} \paq{3(F\sdot\phi)^2(F\sdot W_1)(F\sdot
  W_2)+ \nonumber \\ 
  && -\frac{N_c}{2}f_{abc}W_1^b\phi_cf_{ade}W_2^d\phi_e}-(1\leftrightarrow 2)}
\end{eqnarray}

Introducing a Cartan-Weyl basis for $SU(N_c)$, $S^{(1)}(\mu)$ can be
written as
\begin{eqnarray}\label{cw}  
  S^{(1)}(\mu) &\sim& 2\pintf\pag{\tr(\www{W}^2)+ \nonumber\\ 
  && -\frac{g_0^2/8}{(p^2+\mu)^2} \paq{3\sum_\alpha(F\sdot 
  \phi)^2_{\alpha\alpha}(\vec{\alpha}\sdot\vec{W})^2
  -N_c\sum_\alpha(\vec{\alpha}\sdot\vec{W})^2\phi_{-\alpha}\phi_{\alpha}}}
\end{eqnarray}
where $\vec{\alpha}$ refers to the roots and $\vec{W}$ is taken to
belong to the Cartan subalgebra.

In our low energy (potential) approximation only the charged
components of the chiral scalar contributes to the effective action,
so one may replace $(F\sdot\phi)_{\alpha\alpha}^2$ by
$(E_\beta\phi_\beta)^2_{\alpha\alpha}$, where $E_\alpha$ are the ladder
operators in the $SU(N_c)$ algebra. Then \be
\sum_\alpha(F\sdot\phi)^2_{\alpha\alpha}(\vec{\alpha}\sdot\vec{W})^2
=
\sum_\alpha(E_\beta\phi_\beta)^2_{\alpha\alpha}(\vec{\alpha}\sdot\vec{W})^2=
\sum_{i>j}\ppq{N_c(W_i^2+W_j^2)+W^2}\phi_{\mu_i-\mu_j}\phi_{\mu_j-\mu_i}
\ee where $\mu_i$ ($i=1,...,N_c$) are the weights of the fundamental
representation of $SU(N_c)$.

As for the second term (the commutator term) \be
-N_c\sum_\alpha(\vec{\alpha}\sdot\vec{W})^2\phi_{-\alpha}\phi_{\alpha}
=
-N_c\sum_{i>j}\ppq{(W_i-W_j)^2}\phi_{\mu_i-\mu_j}\phi_{\mu_j-\mu_i}\ee
Eq. (\ref{cw}) becomes 
\begin{eqnarray} 
  S^{(1)}(\mu)&=&2\pintf\pag{\tr(\www{W}^2)-\frac{g_0^2/8}{(p^2+\mu)^2}\cdot
  \nonumber\\ 
  &&\cdot\sum_{i<j}\ppq{2N_c(W_i^2+W_j^2+W_iW_j)+3W^2}
  \phi_{\mu_i-\mu_j}\phi_{\mu_j-\mu_i}}.
\end{eqnarray}

Thus $S^{(1)}(M_0)-S^{(1)}(M)$ consists of two terms:
\begin{enumerate}
\item\textbf{Constant ($\fv{3}$ independent) term} \be\label{const_n}
2\pintf\tr\www{W}^2\biggl|^{M_0}_{M}=
\frac{2N_cW^2}{16\pi^2}\intg\frac{\tau\di{}{\tau}}{(\tau+\mu)^2}\biggl|^{M_0}_{M}
= \frac{N_cW^2}{16\pi^2}\log\pat{\frac{M}{M_0}}^2\ee
\item\textbf{$\fv{3}$ mass term}
\bea\pint\frac{2g_0^2/8}{(p^2+\mu)^4}\sum_{i<j}\ppq{2N_c\omega_{ij}+3W^2}
\phi_{\mu_i-\mu_j}\phi_{\mu_j-\mu_i}\biggl|^{\mu=M_0}_{\mu=M}\nonumber\\
=\frac{2g_0^2}{16\pi^2 8\cdot
6M^2}\sum_{i<j}(2N_c\omega_{ij}+3W^2)\phi_{\mu_i-\mu_j}\phi_{\mu_j-\mu_i}
\eea where $\omega_{ij}\equiv W_i^2+W_j^2+W_iW_j$ and we have
omitted the term proportional to $1/M_0^2$.
\end{enumerate} Thus after integrating $\fv{1}$ and $\fv{2}$, the
effective action for $\fv{3}=\phi$ is, up to the quadratic term,
\be\label{eff-coupling-n}
\ptint\frac{1}{2}\sum_{i>j}\pag{-p^2+\pi\www{W}+\paq{\mu+\frac{2}{32\pi^2}\frac{g_0^2}{8}\frac{1}{3M^2}\pat{2N_c\omega_{ij}+3W^2}}}\phi_{\mu_i-\mu_j}\phi_{\mu_j-\mu_i}\ee

This corresponds to eq. (\ref{eff-coupling}) of section \ref{chap_2_qft} in the
case of $N_c=2$. Note however that, due to the different treatment
with respect to the anticommuting external field $\vec{W}^{(\alpha)}$, eq.
(\ref{eff-coupling-n}) would not go over formally to eq.
(\ref{eff-coupling}) simply with $N_c=2$ and $W_1\sdot W_2=W^2/2$ (see
below).

To obtain the final form of the effective potential one integrates over 
$\fv{3}$ with the approximate quadratic action (\ref{eff-coupling-n}).

Here one has again to apply the RG transformation to separate the
contribution for $\N=4$ SYM $S^{(2)}(M)$ ($M\to0$) and \virg{Wilsonian
part}, $S^{(2)}(M_0)-S^{(2)}(M)$. Because of the gaussian
approximation, one can repeat the same arguments used for
integrating out $\fv{1}$ and $\fv{2}$, making use of an
appropriate rescaling transformation.

The resulting non zero contribution to the effective action, $S^{(2)}$
(function of $\vec{W}$ only), is \bea
S^{(2)}(\mu)&=&\frac{1}{4}\ptint\sum_{i\neq
j}\pi^\alpha(W_i-W_j)_\alpha\pi^\beta(W_i-W_j)_\beta\times \nonumber\\
&&\times\paq{p^2+\mu+\frac{g_0^2}{64\pi^2\cdot3M^2}(3W^2+2N_c\omega_{ij})}^{-2}\nonumber\\
&=&\frac{1}{8}\pint\sum_{i,j}(W_i^2+W_j^2-W_iW_j-W_jW_i)\times\nonumber\\
&&\times\paq{p^2+\mu
+\frac{g_0^2W^2}{64\pi^2 M^2}+\frac{g_0^2N_c\omega_{ij}}{32\pi^2 \cdot 3 
M^2}}^{-2} 
\eea 
If one expands the last expression in powers of
$\omega_{ij}$, then the generic term is of the form
\bea\label{s_final_n}
S_n&\equiv&\const\pint\sum_{i,j}(W_i^2+W_j^2-W_iW_j-W_jW_i)\paq{p^2+\mu
+\frac{g_0^2W^2}{64\pi^2 M^2}}^{-2}\times\nn
&&\times(-)^n (n+1) \paq{\frac{\frac{g_0^2W^2}{32\pi^2\cdot 3M^2}}{p^2+\mu+
\frac{g_0^2W^2}{64\pi^2 M^2}}}^{n}\omega_{ij}^n
\qquad (n=0,1,2,...)  \eea 
Here one must remember that $W^\alpha_i,W^\beta_j$ are
anticommuting fields. One consequence of this is of course that \be
(W^2)^\nu=0\quad\text{for}\quad\nu\geq N_c\ee However, let us assume
that $N_c$ is sufficiently large so that one can attach unambiguous
meaning to the function defined by the series
\be\sum_{\nu=0}^{\infty}C_\nu(W^2)^\nu\ee to any desired order.

The important point is the fact that in (\ref{s_final_n}), for any
given pair $i\neq j$, one cannot have the product of more than
four $W_i$ and $W_j$, \ie $S_n=0$ except for $n=0$ or $n=1$.

Then eq. (\ref{s_final_n}) reduces to \bea
S^{(2)}(\mu)&=&\frac{1}{8}\pint\sum_{i,j}(W_i^2+W_j^2-W_iW_j-W_jW_i)\times\nn
&&\times\paq{p^2+\mu+\frac{g_0^2W^2}{64\pi^2
    M^2}}^{-2}\Biggl(1-\frac{\frac{2g_0^2}{32\pi^2}\frac{N_c
    \omega_{ij}}{3M^2}}{p^2+\mu+\frac{g_0^2W^2}{64\pi^2 M^2}}\Biggr)\nn
&=&\frac{N_c}{8}\pint\paq{p^2+\mu+\frac{g_0^2W^2}{64\pi^2 M^2}}^{-2}
\Biggl(W^2-\frac{\frac{g_0^2}{32\pi^2}\frac{(W^2)^2}{M^2}}{p^2+\mu+\frac{g_0^2W^2}{64\pi^2
M^2}}\Biggr)\eea

Note that, to arrive at the above, one has used the
completeness relation for the $SU(N_c)$ weights 
\be\sum_{i=1}^{N_c}\mu_i^A\mu_i^B=\delta_{AB}/2\ee

Thus, the superpotential can be determined from 
\bea\label{sup_n}
S^{(2)}(M_0)-S^{(2)}(M)&=&\frac{N_c}{8\cdot
16\pi^2}\!\!\int_0^\infty\!\!\frac{\tau\di{}{\tau}}{\ppq{\tau+\mu+\frac{g_0^2W^2}{64\pi^2
M^2}}^2}\Biggl[W^2-\frac{\frac{g_0^2(W^2)^2}{32\pi^2
M^2}}{\tau+\mu+\frac{g_0^2W^2}{64\pi^2
M^2}}\Biggr] \Biggr|^{\mu=M_0}_{\mu=M}\nn &=&\frac{N_cW^2}{8\cdot
16\pi^2}\Biggl\{\log\Biggl(\frac{M+\frac{g_0^2W^2}{64\pi^2
M^2}}{M_0+\frac{g_0^2W^2}{64\pi^2
M^2}}\Biggr)+\frac{1}{2}\frac{\frac{g_0^2W^2}{32\pi^2
M^2}}{M+\frac{g_0^2W^2}{64\pi^2 M^2}}\Biggr\} \eea Again one has
omitted $1/M_0^2$ term.

Naturally, eq. (\ref{sup_n}) will \textbf{not} go over to eq. (\ref{sup_2}) of 
section \ref{chap_2_qft} when $N_c=2$. This is, as mentioned above, due to
the different treatment of the anticommuting fields. Roughly speaking, 
eq. (\ref{sup_2})
makes unambiguous sense only up to $\mathcal{O}(W^2)$.

\subsection{IR limit of superpotential and VY form}

According to the assumptions discussed in the Introduction, eq. (\ref{sup_n}) 
should lead to the superpotential for $\N=1$ SYM (or more generally, to the
perturbative expansion of $\N=1^*$ model with small parameter
$g_0^2W^2/M_0^3$) when $M\to 0$, that is when the residual terms should 
reproduce the superpotential for $\N=4$ SYM, \ie a triviality.

The second term in the last line of eq. (\ref{sup_n}) will go over to
constant as $M\to 0$ (or, more appropriately, as $\sim M^3/W^2\to 0$), so we
may substitute its contribution for small $M$ by \be \frac{N_c}{8\cdot
16\pi^2}W^2\ee

To the one loop integral (\ref{sup_n}), one must add the
constant term (\ref{const_n}) and the kinematical term in $S_{\N=4}$
SYM which is equal to \be \frac{1}{16\pi^2}\frac{1}{g_0^2}\intg
W^2=\frac{1}{16\pi^2}\frac{N_c}{8}\pag{\log\paq{\pat{\frac{M_0}{\Lambda}}^3\frac{1}{g_0^2}}+\frac{\imath\vartheta_0}{N_c}}\intg
W^2\ee

The final result is then \be\label{total_n}
\frac{N_c}{128\pi^2}\intg W^2\Biggl\{
\log\paq{\pat{\frac{M}{\Lambda}}^3\frac{1}{g_0^2}}+
\log\Biggl(\frac{1+\frac{g_0^2W^2}{64\pi^2
M^3}}{1+\frac{g_0^2W^2}{64\pi^2 M_0M^2}}\Biggr) + 1 + \frac{\imath
\vartheta_0}{N_c}\Biggr\}\ee

To go further and deduce the VY form of the potential, one has to
follow the same argument as in section \ref{chap_2_qft}~\cite{7} 
[cfr.~eq.~(\ref{const_n})]: if one is allowed to conclude that the second,
logarithmic term in eq. (\ref{total_n}) can be replaced for \virg{small $M$} by
\be\sim\log\pat{\frac{g_0^2W^2}{64\pi^2 M^3}}, \ee
then the whole of eq. (\ref{total_n}) can be reduced to VY form: \be\label{97}\sim
\frac{N_c}{128\pi^2}\intg W^2\pag{\log\paq{\frac{W^2}{128\pi^2}\biggl/\frac{2 
\Lambda^3 }{\e}} +\frac{\imath\vartheta_0}{N_c}}.\ee 

Eq. (\ref{97}) should be
compared with the standard expression \be\frac{N_c}{128\pi^2}\intg
W^2\pag{\log\paq{\frac{W^2}{128\pi^2}\biggl/\Lambda^3\e}
+\frac{\imath\vartheta_0}{N_c}}\ee

Now, to assert
\be\label{limits}\log\paq{\frac{1+\overline{\alpha}
g_0^2W^2/M^3}{1+\overline{\alpha} g_0^2W^2/(M_0M^2)}}
\sim\log\ppt{\overline{\alpha} g_0^2W^2/M^3}\ee ($M_0\gg M$,$M\sim 0$,
$\overline{\alpha}=$ numerical constant) it is necessary and
sufficient to have \be\label{3_21} \overline{\alpha} g_0^2W^2/M^3\gg 1
\quad \text{and}\quad  \overline{\alpha}
g_0^2W^2/(M_0M^2)\equiv(M/M_0)\overline{\alpha}g_0^2W^2/M^3\ll 1 \ee

Substituting to $W^2$ its \virg{desired} mean value
$W^2\sim\Lambda^3$, (\ref{limits}) becomes \be g_0^2(\Lambda/M)^3\gg
1\quad \text{and}\quad (M/M_0) g_0^2(\Lambda/M)^3 \ll 1 \ee or \be
(M_0/M) \gg g_0^2(\Lambda/M)^3 \gg 1 \ee

By the definition of the dynamical cutoff of $\N=1$ SYM \ben
g_0^2(\Lambda/M)^3=(M_0/M)^3\exp{\ppq{-8\pi^2/(N_cg_0^2)}} \een In the
end, one needs the simultaneous inequalities \ben
(M_0/M)^2\exp{\ppq{-8\pi^2/(N_cg_0^2)}}\ll 1 \quad \text{and}\quad
(M_0/M)^3\exp{\ppq{-8\pi^2/(N_cg_0^2)}}\gg 1\een or \be\label{ineq}
\exp{\frac{1}{3}\ppq{8\pi^2/(N_cg_0^2)}}\ll (M_0/M) \ll
\exp{\frac{1}{2}\ppq{8\pi^2/(N_cg_0^2)}} \ee Does this last set of
inequalities make sense for values of $M_0/M$ between $1$ and $\infty$?

Naturally, this makes sense only for \be\label{cond}
\exp{\ppq{8\pi^2/(N_cg_0^2)}} \gg 1.\ee

Only then $(M/M_0)$ can flow into $M/M_0\to 0$. (\ref{cond}) implies
the 't Hooft coupling should be small \ben \lambda^2=N_cg_0^2\to 0\een

On the other hand, in the strong coupling regime, \be
\exp{\ppq{8\pi^2/(N_cg_0^2)}} \lesssim 1, \ee
the inequalities
(\ref{ineq}) do not make any sense ($M/M_0\sim 1$) and the present
computational scheme collapses.

\subsection{Non-Gaussian corrections}

The above discussion is relevant also for another difficulty
raised in section \ref{chap_2_qft}, \ie the justification of the gaussian
approximation.

The full effective action for $\fv{3}=\phi$, after 
$\fv{1}$ and $\fv{2}$ have been integrated out, looks like \be
\label{non_gaussian}\intg\pat{\frac{1}{2}\vec{\phi}\ppt{\www{\Gamma}+M_0}\vec{\phi}+W^2\zeta'\ppt{\frac{g_0}{M}\phi}}
\ee where $\www{\Gamma}=-p^2+\pi\www{W}+(1/32\pi^2)(g_0^2/3M^2)W^2$.

To simplify the discussion, here we have taken the special case of
$N_c=2$, the generalization to arbitrary $N_c$ being straightforward.

The rescaling transformation, which can be written as \ben \phi
\to \sqrt{\frac{\www{\Gamma}+M}{\www{\Gamma}+M_0}}\phi \een transforms eq.
(\ref{non_gaussian}) to \be\label{action_resc}
\intg\pag{\frac{1}{2}\vec{\phi}\ppt{\www{\Gamma}+M_0}\vec{\phi}+
W^2\zeta'\pat{\frac{g_0}{M}\sqrt{\frac{\www{\Gamma}+M}{\www{\Gamma}+M_0}}\phi}}
\ee

From eq. (\ref{action_resc}), one can read off the components of relevant
Feynman graphs.

An internal line connecting any two vertices in $\zeta'$ looks like
\ben \sim
\pat{\frac{g_0}{M}}^2\frac{1}{\www{\Gamma}+M_0}. \een

Thus, for a vacuum graph with $I$ internal lines, $V$ vertices and $L$
loops, its value is proportional to \bea F(I,V,L)&\propto&
S^LS^V\paq{\pat{\frac{g_0}{M}^2}^2\frac{1}{\www{\Gamma}+M_0}}^I
=S\pat{\frac{S(g_0/M)^2}{\www{\Gamma}+M_0}}^I\text{ since
$L+V=I-1$}\nn &\sim&
S\Biggl(\frac{g_0^2\frac{W^2}{M^2}}{M_0+\frac{g_0^2W^2}{32\pi^2\cdot
3M^2}}\Biggr)^I=
S\Biggl(\frac{g_0^2\frac{W^2}{M_0M^2}}{1+\frac{g_0^2W^2}{32\pi^2\cdot
3M_0M^2}}\Biggr)^I \eea

Thus, again, the error due to the gaussian approximation is negligible
only if (cfr. eq. (\ref{3_21})) \be \frac{g_0^2W^2}{M_0M^2}\ll 1.\ee

\section{Conclusions}\label{chap_4}

In this note, we have attempted to explain the VY potential of $\N=1$
SYM model with gauge group $SU(N_c)$ starting with the microscopic
Lagrangian and covariant supersymmetric (\virg{holomorphic}) Feynman
rules, valid for low energy external states.

Instead of the usual instanton expansion, we have applied a Renormalization Group-inspired 
method of varying the regularizing mass $\mu$ ($M_0\geq\mu\geq
M$). Taking the limit that $M\to 0$, one hopes to deduce the potential
in question as the difference with respect to the holomorphic
superpotential of $\N=4$ SYM model which is assumed to be trivial.

In the end, we have obtained, with a more or less reliable
approximation, the superpotential of pure $\N=1$ SYM. Note that this
is obtained essentially as one loop effect, as has been stated or
conjectured several times previously \cite{1}.

Indeed, the RG method allows one to extract the convergent expression
for such a one loop integral, which can then lead to VY form for the case 
of pure SYM.

However, we have to conclude also that our method, while qualitatively
correct, does not arrive at the precision and generality of the Matrix
Model approach. As we have seen in (\ref{chap_2}), the Matrix Model
can be applied to much wider class of problems.

Once one accepts the prescription of \cite{3} with the fixed
integration measure of \cite{10} one can obtain the superpotential
of $\N=1$ and $\N=2$ models without any ambiguity.

On the other hand, in \cite{13} the
\textbf{direct} correspondence (without going over to the superstring
theory or M-theory) has been shown between supersymmetric gauge field theory
and matrix model.

More explicitly, one can map the $\N=1$ $U(N)$ SYM + adjoint matter on
non-commutative space time on to the large $\hN$ limit of certain
super matrix models.

Since the former at low energies goes over to the usual GFT model on
commutative space-time except for the quantities with UV
divergence, the authors of \cite{13} apply this correspondence to
derive directly the Dijkgraaf-Vafa method.

In the present note, the main object of the discussion is the so-called
$\N=1^*$ model - mass deformed $\N=4$ SYM - which is free of UV
divergences. Thus the demonstration of Kawai et al. must apply in a
very simple way.

Then, there \textbf{must} be a simple QFT method which reproduces
the Matrix Model results for quantities like superpotentials.

One possibility is that our present method does not take sufficient
account of the IR structure of the $\N=4$ model. Indeed, we did not find the
\virg{miracle} corresponding to the large $\hN$ matrix result quoted
in section \ref{chap_3}.

At the same time, the problem of singular external fields like $S\equiv
W^2$ with $S^{N_c}=0$ ($S$ is however \virg{bosonic}) appears not to
have been completely cleared. In fact, one must recognize that the
discussion on IR limit at the end of  section \ref{chap_3} is still not
entirely satisfactory. For instance, for the simplest case of $N_c=2$,
one can even invent a \virg{proof} of the desired IR limit.
Namely, by making use of the fact that $(W^2)^2 \equiv 0$ for $N_c = 2$, one
can easily show that
\bean
&&\log\pag{\frac{1+A/M^3}{1+A/(M_0M^2)}}+\log\pat{\frac{M}{\Lambda}}^3\\
&\approx&
\pat{1-\frac{M}{M_0}}\pag{\log\pat{1+\frac{A}{M^3}}+\log\pat{\frac{M}{\Lambda}}^3}+\frac{M}{M_0}\log\pat{\frac{M}{\Lambda}}^3,
\eean
where $A\propto W^2$.
The last expression has a finite $M \to 0$ limit which is
precisely equal to $\log(A/\Lambda^3)$ as desired.

This \virg{demonstration} only shows that we did not yet establish the
exact rule for dealing with \virg{classical} quantities like
$S\equiv W^2$ and their functions, quite apart from the quantum effect
discussed in \cite{16}.  These problems need further investigations.

\section*{Acknowledgements}

Part of the work has been done while KY was staying at Riken (Wako-Shi, 
Japan) and YITP (Kyoto, Japan). He is grateful to the warm hospitality tended
to him by T.~Tada (Riken) and T.~Kugo and M.~Ninomiya (YITP).

We have profited from numeorus discussion with colleagues, in particular 
F.~Guerrieri, H.~Kawai, K.~Ohta, G.C.~Rossi, Ya.S.~Stanev, and M.~Testa.

Special thanks are due to T. Morita for numerous instructive discussions on
Matrix Model as well as for communicating the results contained in his thesis
prior to their publications.


\end{document}